# Memory of past designs:
# distinctive roles in individual and collective design

Françoise Détienne
EIFFEL Research Group "Cognition and Cooperation in Design"
INRIA
Domaine de Voluceau,
Rocquencourt, BP 105,
78153, Le Chesnay, France
Email: Francoise.Detienne@inria.fr

**Abstract**

Empirical studies on design have emphasised the role of memory of past solutions. Design involves the use of generic knowledge as well as episodic knowledge about past designs for analogous problems : in this way, it involves the reuse of past designs. We analyse this mechanism of reuse from a socio-cognitive viewpoint. According to a purely cognitive approach, reuse involves cognitive mechanisms linked to the problem solving activity itself. Our socio-cognitive approach accounts for these phenomena as well as reuse mechanisms linked to cooperation, in particular coordination, and confrontation/integration of viewpoints.

Keywords: design, reuse, team work

## 1    Design problems: some characteristics

### 1.1    "ill-defined" problems

Cognitive ergonomics does not identify design in relation to a social function or a status, but qualifies as design tasks certain professional activities in which a set of formal characteristics can be identified. Therefore, one can identify numerous professional domains that deal with design. It can be the design of material artefacts (e.g. mechanical engineering, electronics, architecture) or the generation of symbolic or abstract devices (e.g. planning or computer programming).

The most common conception of design problems considers them as "ill-structured" problems (Eastman, 1969; Falzon et al., 1990; Simon, 1973; Visser & Hoc, 1990). Their characteristics are as follows:
- the specifications given at the start are never complete or unambiguous: initial problem specifications are not sufficient to define the goal, i.e., the solution, and stepwise definition of new constraints is necessary;
- the resolution of conflicting constraints, often coming from different representations and processing systems, plays an important role;
- there is no definite criterion for testing any proposed solution, as is typically the case for "well-structured" problems: design problem solutions are more or less "acceptable" or satisfactory, they are not either "correct" or "incorrect";
- various design solutions are acceptable, one being more satisfactory according to one criterion, another according to another criterion.

- Problems tend to be large and complex. They are not generally confined to local problems, and the variables and their interrelations are too numerous to be divided into independent sub-systems.

One consequence of this complexity is that the resolution of these problems often requires that multiple skills be put together, which leads to development of collaboration within a single working group.

### 1.2  A collective process

In individual problem-solving, the division of problems into sub-problems results in a reduction in complexity. In team work, one consequence of solving a complex problem is that the solution requires a combination of various skills, which in turn leads to the development of cooperation processes between individuals. Developing and maintaining such cooperation may be seen as an additional task in team design.

For the distributed decision-making system, tasks are distributed among design teams, each carrying out various sub-tasks. Beguin (1997) highlights that, as soon as tasks are divided, conflicts and negotiations between designers arise. Solutions are therefore not only acceptable in terms of problem-solving itself. They result from some compromise between designers: solutions are negotiated.

Team design can be characterised as cycles of distributed design and co-design phases. In the distributed design phase, the actors who are simultaneously (but individually) involved in the same co-operation process, carry out well-determined tasks. They pursue goals (or at least sub-goals) that are specific to them. In the co-design phase, actors share an identical goal and contribute in order to reach it through their specific skills. They do so with very strong constraints of direct co-operation so as to guarantee a solution to the problem resolution. These two design phases involve distinct co-operation processes (Falzon, 1994; Falzon & Darses, 1996):
- Operative synchronisation and co-ordination in distributed design;
- Cognitive synchronisation and confrontation/integration of viewpoints in co-design.

Operative synchronisation is crucial in distributed design. It fulfils two functions. Firstly, it aims at ensuring that the tasks are shared between the partners of the team activity. Secondly, it aims at ensuring, the start, the end, the simultaneity, the sequencing, and the rhythm of the actions to be carried out. Operative synchronisation leads to co-ordination activities. Grinter (1999) describes two mechanisms to facilitate cross-group co-ordination in distributed design : boundary spanners and boundary object. She identifies :
- Boundary spanners as people who move among different teams transferring information about the state of the project. They translate information from a form given by one team into a form that could be understood by other teams. Boundary spanners are characterised as an informal role, adopted by persons with good communication skills who have contacts with various teams. They are often essential in the communication between rival teams .
- Boundary objects are objects adapted to the local needs and constraints used by various teams and shared by all the actors of the project.

Cognitive synchronisation allows the participants:

- To ensure that they share knowledge about the state of the situation: e.g., problem data, state of the solution.
- To ensure that they share the same general knowledge about the domain: e.g., technical rules, domain objects, solving procedures.

Thus the objective of cognitive synchronisation is to construct a common operative referential (de Terssac & Chabaud, 1990). Confrontation and integration of viewpoints consists in negotiation and argumentation about solutions produced by the various co-designers in the objective to converge toward common options of design.

### 1.3   Reuse of past designs as a cognitive invariant

There are cognitive invariants in the design activity whatever the application domain: e.g reuse of past designs. Indeed, the design of an artefact is based on the use of generic knowledge, but also, on the use of episodic knowledge about an analogous design situation dealt with in the past (Visser, 1995, Visser & Trousse, 1993). In this case, the designers may or may not have an external representation (e.g. design rationale) about this past design at their disposal.

Analogical reasoning models (e.g., Clement, 1998; Gentner, 1989) offer a theoretical framework suitable for studying the cognitive mechanisms of reuse. Two major phases are distinguished in the study of reasoning by analogy: access and use. In the access phase, a source situation is selected or retrieved from memory. In the use phase, the source situation is applied to the current target situation. More precisely, we can distinguish:
-   the construction of a representation of the target situation;
-   retrieval or selection of a source situation analogous to the target situation;
-   mapping between the source situation and the target situation;
-   adaptation of the source solution to solve the target problem.

The reuse mechanism has mostly been analysed in the purely cognitive approach (e.g., Maiden, 1991; Sutcliffe & Maiden, 1991). In this paper, we intend to revisit this mechanism of reuse from a socio-cognitive approach. According to a purely cognitive approach, reuse involves cognitive mechanisms linked to the problem solving activity itself and to the individual design activity. Our socio-cognitive approach accounts for these phenomena as well as reuse mechanisms linked to cooperation in collective design, in particular coordination, and confrontation/integration of viewpoints.

### 2   Toward a socio-cognitive classification of reuse situations

Our research question is to construct a classification of reuse situations characterised by distinctive processes involved in reuse. One classification that is currently used in engineering is based on the type of component reused, which is associated with a preferred type of activity. For example, Software Engineering distinguishes the extraction of code from an existing application (for example, lines of codes or procedures), specialisation of components, generally taken from a library, and finally inheritance/composition of classes. For each of these situations, reuse is associated with modifications at different levels.

We have identified three other dimensions which can be used to construct a cognitive classification of reuse situations:

- Prospective versus retrospective reuse: this concerns the temporal management of the organisation of the design activity.
- Reuse in planning versus reuse in translating: this concerns the level of representation, abstract versus detailed , and the level of control of the design activity.
- Reuse as a problem solving mechanism versus reuse as an argumentation mechanism: this concerns the kind of process involved, either a cognitive process or an interactional process, more particularly, an argumentation process.

This cognitive classification will allow us to identify which cognitive mechanisms are specific to certain reuse situations. For example we will see that:
- mechanisms of anticipation belong to the situation of prospective reuse ;
- mechanisms for enriching the representation of the target are specific to reuse in planning whereas the lowering of the level of control of the activity is specific to reuse in translating.

From a cognitive ergonomics viewpoint, this socio-cognitive analysis of the reuse situation will allow us to identify which support is needed depending on the reuse situation. Our cognitive classification of reuse situations allows us to suggest different reuse aids.  Different types of reuse episode require different types of support. After having rapidly referred to our empirical basis , we will develop for each dimension, which processes are involved and which supports are expected for these processes.

## 2.1     Empirical basis

Our theoretical development is based on results of empirical studies conducted by the author and of other empirical studies in the literature. The characteristics of the most significant of these studies are displayed in Table 1. We have highlighted some key characteristics of these studies:
- Application domains and design environments: the application domains are varied: software engineering, aeronautical engineering, architecture. This will support our view of invariant design characteristics, in particular the reuse of past designs, whatever the application field. Furthermore, we could argue that some application domains, like software engineering, are more dedicated to reuse because specific characteristics of the environments support it.  The studies in other application domains where such supports do not exist show that these kinds of mechanisms are also involved.
- Type of studies and participants: there are field studies or experiments. Classical laboratory studies in cognitive psychology (see for example, Gentner, 1989) show that the use of analogical reasoning is not, in general, spontaneous. One limitation of these studies is that the subjects are seldom experts in the application domain. Here we focus on situations in which expertise in an application domain is a major characteristic of the subjects. In this condition, the reuse of past designs through analogical reasoning is a spontaneous process in field studies, in which the designers are in familiar design problem situations, as well as in experiments, in which the design task and environment retain some essential features of their usual tasks.
- Collective versus individual design: the studies concern individual design as well as collective design. It was essential to have this broad spectrum in order to reach our goal of analysing the distinctive role of memory of past design in individual and collective design.
- Collected data: the focus of the studies is on the design process, through the analysis of verbalization, dialogues and documents processed, rather than on the design product. This is an essential feature of the studies which allow the authors to analyse in a detailed way the strategies used in design. These analyses and their results are the basis of our theoretical framework.

| Study's references | Application domain | Type of study | Participants | Collective versus individual design | Design environment | Collected data |
|---|---|---|---|---|---|---|
| Détienne, 1991 | Software | Experiment | 4 professional programmers | Individual design | Object-oriented language and environment | Verbalization, successive versions of programs, notes |
| Burkhardt & Détienne, 1995 | Software | Experiment | 7 professional programmers | Individual design | Object-oriented language + paper/pencil | Verbalization, successive versions of programs, notes |
| Martin et al. 2000a | Aeronautic | Field study | Professional designers in 8 fields | Collective design/ distributed design phases | CAD PDM | Interviews, observations in the design office |
| Martin et al., 2000b; 2001 | Aeronautic | Field study | Professional designers in 8 fields | Collective design/ Co-design phases | CAD PDM | Dialogues and data produced and consulted during (7) meetings + interviews afterwards |
| Rosson & Carroll, 1993 | Software | Experiment | 4 professional programmers | Individual design | Object-oriented language and environment | Verbalization, successive versions of programs, notes |
| De Vries, 1993; 1994 | Architecture | Experiment | 51 Students in architecture | Individual design | Paper/pencil + information system of reusable designs | Exploration and use of the information system |
| Visser, 1987 | Software | Field study | 1 professional designer | Individual design | Declarative boolean language | Verbalization, successive versions of programs, notes |

Table 1. Some key characteristics of studies of design in which reuse is involved

## 2.2   Dimension 1: Prospective reuse versus retrospective reuse

One way of classifying reuse depends on whether the reuse episode begins with the source development, prospective reuse, and if not, retrospective reuse. Figure 1 distinguishes these two situations, on a temporal axis of design.

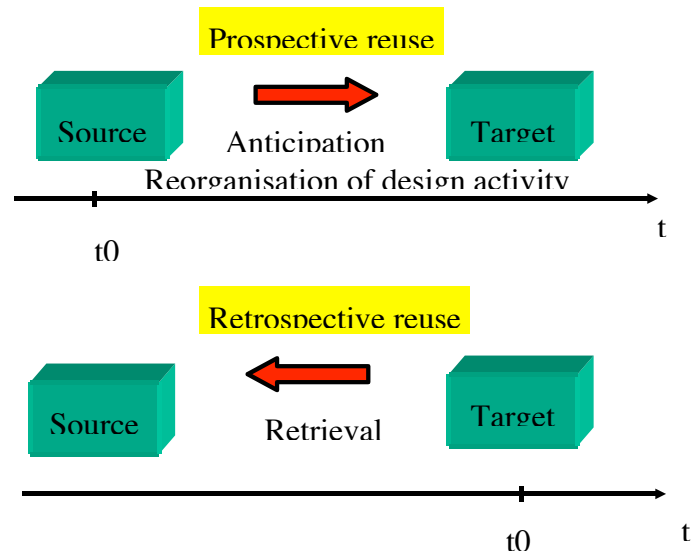

Figure 1: Prospective *versus* retrospective reuse

### 2.2.1 Retrospective reuse

When retrospective reuse (Burkhardt & Détienne, 1995; Détienne & Burkhardt, 2001) is involved, the designer retrieves a source-solution previously developed and adapts it for a current problem, the target. This situation, quite classically described in the literature, implies that the mechanisms used relate to the retrieval of a source and the exploration of the relationship between the source and the target, in order to develop the new solution.

### 2.2.2 Prospective reuse

In prospective reuse (Détienne, 1991; 2002), the designer develops a solution to a sub-problem (which will have the status of the source) and envisages reusing it to solve other sub-problems that are still to come (which will have the status of targets). The source and the target will represent instances of the same schema. A cognitive process peculiar to this situation is the process of looking ahead while developing the source. In effect, the designer anticipates the way that the source will be adapted to construct the target solutions. This mechanism manifests itself in two forms:
- construction of an operative representation of the source, that is, one which distinguishes clearly between the fixed features of the source and the features that may be varied when it is reused;
- construction of an adaptation procedure from the source to the target(s); this adaptation procedure allows the variable parts to be modified in order to create other instances of the same schema.

Another cognitive mechanism peculiar to this situation is reorganization of the design activity. If, in the original plan, different elements that are considered to be instances of a single schema are scheduled to be developed at widely separated times during the design, the designer will reorganize the plan in order to develop these instances in a row. This way, it entails opportunistic deviations from the original plan (Guindon, 1990; Visser, 1987). This allows the representation of the source and the adaptation procedure to be kept in working memory, without adding representations constructed in the course of other development activities. It thus minimizes the risk of errors of omission.

2.2.3   Implications

For retrospective reuse, support for the retrieval and understanding of source situations needs to be considered. Support to retrieval is quite a classical focus in research on reuse tools. Support to understanding has been rather neglected. As specifying this kind of support depends on dimension 2, we will develop this point in the next section.

For prospective reuse, it would be useful to have support for the visualisation of the operative representation of the source and the construction of the modification procedure, constructed in this situation.  Such assistance could form the basis for the automatic generation of target systems.  Further, it would be also helpful to support the opportunistic organization of the design activity as this kind of reuse episode triggers opportunistic deviations from the original plan. Recommendations on support of these kinds have been made in the literature (see Guindon 1992).

**2.3   Dimension 2: Reuse in planning versus reuse in translating**

In design, we can identify cycles of planning and translating. One of the most influential cognitive models of text production is that proposed by Hayes and Flower (1980). Hayes and Flower have defined three major phases in the writing process: Planning of the text structure as a function of domain knowledge (organizing) and communication purposes (goal setting); translating the text plan into a linguistic representation; and reviewing the text as a function of the writer's evaluation. One important feature of this model is that the overall process is cyclical rather than strictly linear (Bereiter et al. 1988; Flower & Hayes, 1981).

Design also includes phases of planning, translation and revision, usually called problem solving or design, implementing, revising (Détienne et al., 1996). For example, Gray and Anderson (1987) showed that such cycles occur in software development. Planning involves both retrieving problem-relevant knowledge and building up an abstract solution. Translating is equivalent to implementing the solution in a particular language. Finally, revising may include either modifying the implementation, the abstract solution, or even one's understanding of the problem structure.

Reuse may occur in planning or in translating: in these two situations, a different representation of the source is constructed, as shown in Figure 2.

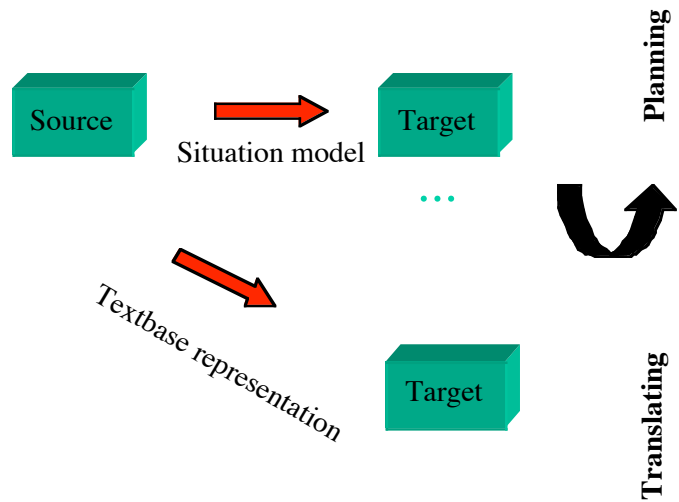

Figure 2: reuse in planning *versus* reuse in translating

### 2.3.1   Reuse in planning in individual reasoning

Reuse in planning has usually been described in individual design. The effect of the reuse processes may be an enrichment of the representations constructed during planning. In software design, Burkhardt and Détienne (1995) show that evoking a reusable component may allow the addition of constraints, and new goals. Rosson and Carroll (1993) note that sometimes the borrowed code is not directly reusable itself but rather is used more as a functional specification. In another domain, architectural design, De Vries (1993; 1994) found that exploiting examples of old designs may allow the inference of new constraints for the new design and allow the constraints to be envisaged at a more abstract levels.

Several studies show that when a source is evoked or retrieved during planning (as opposed to translating), information about the source situation from which the source comes is searched or inferred. In software design, Burkhardt and Détienne (1995) observed that this allowed programmers to infer solution goal structure, constraints, evaluation criteria or design rationales. Rosson and Carroll (1993) note the importance of knowledge about an "example application" of a reusable class.  In their field study, Rouet et al. (1995) found that when selecting a reusable component in a library, designers were looking for information on the application from which the component was extracted. This contextual information, which seems to be highly important, is rarely present in the documentation of components because software engineers generally believe that reusable components must be generic and application-independent.

In all these situations, it seems that reusing a component implies constructing a situation model of the source (van Dijk & Kintsh, 1983) and allows the representation constructed for solving the problem on hand to be enriched and the search space to be enlarged.

### 2.3.2 Reuse in planning in collective reasoning

Reuse in planning can also be involved in collective reasoning. In team design, phases of distributed design are particularly sensitive to dependencies between design fields. Coordination is a key cooperation process for distributed design. The existence of dependencies between various fields stresses the importance of coordination between those fields. Therefore, it is necessary for co-designers to communicate on the state of solutions during inter-field meetings or/and to use tools, e.g., the database, to visualise the state of solutions constructed by experts in other fields. It is also important that the planning of tasks allocated to various fields ensures an improved co-ordination. Coordination problems may occur in relation to at least two aspects:
- communication on the state of solutions produced by various fields can be delayed because some field designers do not update the database in which all solutions acceptable at a given time are stored.
- for task planning and, in particular, the breakdown of problems into sub-problems (area of the designed artefact), designers from different fields can attach different levels of priorities to particular sub-problems. This may entail some gap between the areas of the designed artefact, which are processed by the various field experts at a particular time.

In a field study on aeronautical design (Martin et al., 2000a) it was found that, in order to compensate for a lack of co-ordination, designers within one field construct a shared representation of a "hypothetical" state of the solution of the other fields they depend on. This construction was partly based on informal discussions and on the reuse of specifications of past projects.

As regards this mechanism for field1 designers and field2 designers, it was observed that there existed, on the one hand, a problem of co-ordination between these two fields, and, on the other, a gap between their degrees of progress (refinement of their solution).

For a given zone of the designed artefact, the design field1 solution was less advanced than the field2 solution. This was mainly due to the fact that this zone was judged to be an important sub-problem for Field2 designers but not Field1 designers. However, Field2 designers need specifications from Field1 designers. So it was in some way paradoxical to see that Field2 designers were further ahead.

This paradox was explained as follows. Field2 designers had put a gap reduction mechanism into place. These designers had constructed a representation by default, of a hypothetical solution of the solution provided by field 1 designers for this sub-problem. This mechanism was based on:
- informal discussions between field1 designers and field2 designers. These informal discussions took place between expert designers. They are people who moved among different groups transferring information about the state of the project. In this way, we may consider that their role is one of a boundary spanner.
- Reuse of specifications: field2 designers had reused specifications of field1 designers from a past project of which the current project is a variation. This reuse is based on the personal experience that designers have of a previous project in which they were involved. It is also based on visits made to the assembly line of technically similar aircraft currently in production.

2.3.3   Reuse in translating in individual reasoning

Reuse results in the lowering of the level of control of the activity during the translating phase. We refer to the hierarchy of levels of control developed by Rasmussen and Lind (1982). These authors distinguish between automatic activities, activities based on rules, and activities which involve high-level knowledge. The lowering of the level of control of the activity consists in switching from activities which involve high-level knowledge, e.g. problem solving activities, to activities based on rules and automatic activities, e.g. execution of procedures.

The designer has chosen a design solution that he/she tries to implement by reusing a component that can be fitted into his solution, give or take a few modifications. The lowering of the level of control manifests itself in the use of a trial and error strategy .

The use of the copy/edit style attests this effect. It reflects comprehension avoidance of the copied source and use of surface-level features to construct a representation of it (Lange & Moher, 1989; Rosson & Carroll, 1993). For example, in software design, it has been observed that the designers make "probable" modifications and rely heavily on the debugging tools to evaluate the code. A trial and error strategy involves copying and modifying code. Designers tend to reuse code by copying it and making the modifications that they judge the most likely to be suitable. They try to avoid understanding the source code (a strategy known as comprehension avoidance), by depending on surface features to form hypotheses about its functionality. They then rely on the test and debugging tools at their disposal to modify and correct the code in order to adapt it to the target situation. We note that this strategy is encouraged, if not, indeed, dictated by, the tools available: debugging tools are currently much better developed than tools to help understanding, e.g. suitable documentation.

2.3.4   Implications

Reuse at different stages of the development process changes the cognitive status of the element reused and hence calls for different ways of supporting the understanding of the source. Depending on the type of activity in which the designer is engaged, the type of information sought in the source varies and the support needed is not therefore of the same type. On this last point, it seems that offering several types of knowledge about the past design, accessible in different ways, would help these different processes.

It would thus seem that justification for decisions taken during the design of the source would be useful for reuse in planning while knowledge about implementation details would be more useful for reuse in translating.

One limitation of this approach (Karsenty, 1996) , particularly with regard to design rationale, is the difficulty of predicting all the questions about the justification of the design that designers reusing a component in the future might raise. It has been observed that many questions raised during reuse are not dealt with in the documentation because the culture of the designers has evolved between the production of the component and its reuse. Furthermore, writing documents formalising the design rationale requires more from designers than the simple capture of knowledge; rather, it needs abstract reflection on their own design activity.

Finally, we must point out the limits of reusing specifications from previous design projects in order to compensate for coordination problems. Although such reuse may be possible and efficient in the case of innovative design (e.g. further development of an existing product), it is not necessarily the case for creative design. In this case, it is more important to support the coordination process itself in order to avoid problems of coordination.

## 2.4 Dimension 3: Reuse as a problem solving process versus reuse as an argumentation process

In the previous part of this paper, we have examined reuse as involved in the design problem solving process. We will now show how the memory of past designs can serve collective reasoning, in particular when trying to reach an agreement on a negotiated solution.

### 2.4.1 Reuse as an argumentation process

In multi-fields team design, design solutions are not only produced by individuals specialized in a given field. Due to the team nature of the design activity, solutions are negotiated. These processes typically take place in the meetings which bring together specialists with a co-design aim. Different specialities are going to be present, and they are going to have to justify their design choice so they are going to produce arguments. The purpose of these arguments is to provide information to convince the other people of the pertinence and veracity of the information provided in order to tend towards a conclusion that pushes them towards accepting the proposal. When everyone has a joint will to reach agreement, we shall talk about negotiation. Negotiation does not force a person to accept a solution. Dialogue makes it possible to go towards one conclusion rather than another. The conclusion can be a compromise between what each person wants.

Linguists (Perelman & Olbrechts-Tyteca, 1992; Plantin, 1996) distinguish different kinds of arguments: argument by comparison, argument by analogy, argument of authority.
- An argument by comparison compares several objects in order to assess them in relation to each other. Comparisons can be made by opposition, by classification and quantitative classification.
- An argument by analogy highlights a precedent, i.e. it enables the present case to be compared to a typical case proposed as a model.
- An argument of authority is an indisputable argument which is built on a quotation of statements, so it is in no way proof, even if it is presented as such. In general, the proposer's argument is the fact that it has been expressed by a particular authorized person, on whom he relies, or behind whom he hides. Most of the arguments can take the status of argument of authority depending on factors which give a particularly strong weight to the argument

In a field study conducted in the framework of aeronautical design (Martin et al., 2000b; 2001), it was found that argument by comparison, argument by analogy and argument of authority played a particular role in the integration of viewpoints of designers from various fields (as shown in Figure 3).

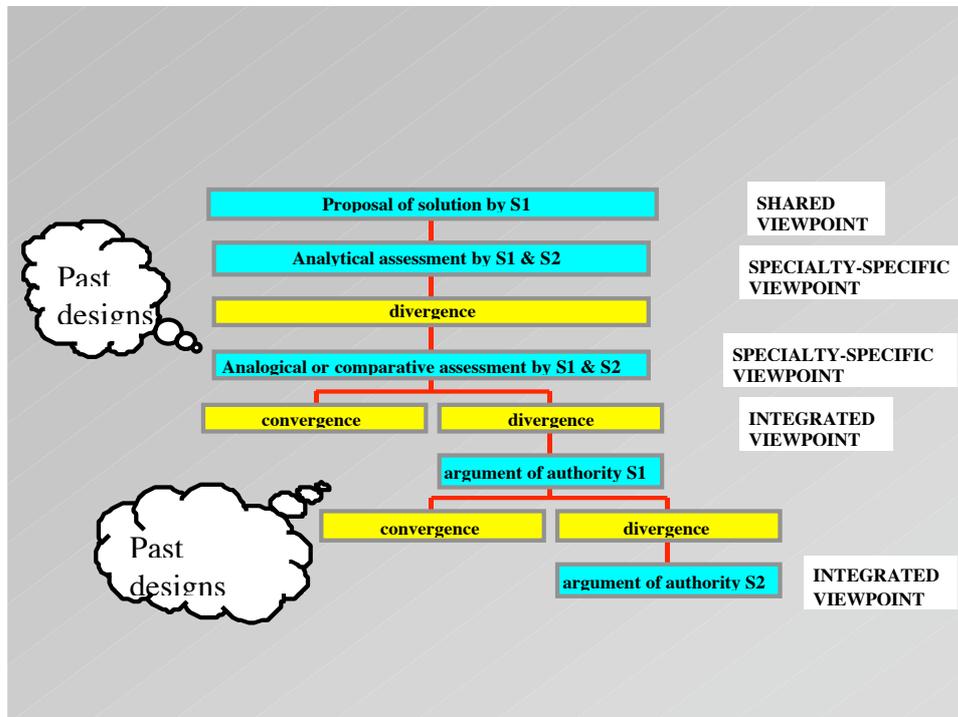

Figure 3: Temporal negotiation pattern in co-design meetings
(S1: speciality 1; S2: speciality 2)

Arguments by analogy served the analogical assessment of the current solution. In this case, there is a transfer of the result of the assessment of an analogical solution (source) developed in the past for the same design project or for a previous design project to the current proposal (target). Whenever the participants shared knowledge about this past design, the argument by analogy was likely to have the strongest weight in the argumentation.

In this case, the shared knowledge about the past design consists in:
- the attributes of the source solution;
- the results of its evaluation process;
- but, most importantly, the various field-dependent constraints used to assess it, the combination and the weighting of these constraints as it was negotiated in the past design: it is the "integrated viewpoint" (Martin et al. 2001) reached by the team in the past.

Evoking attributes of the source and the results of its evaluation is quite classical in analogical reasoning: the distinction here is that it is based on knowledge shared by the team of designer, what is called the common operative referential (Buratto, 2000; De Terssac & Chabaud, 1990). Evoking the "integrated viewpoint" is a mechanism particular to the collective reasoning with its dimensions of expertise and roles.

When knowledge about the past design is not shared, either an argument by authority (relying on the expertise or the status of the person who enunciates it) is involved or traces of the past design process are sought, which takes generally much time.

2.4.2   Implications

In co-design, the shared knowledge of past designs plays an important role, particularly in the negotiation process. It is important to facilitate the construction of this common referential and to ensure the tracability of design decision as well as design rationale. This involves more than applying design rationale techniques (e.g., Questions-Options-Criteria) (see e.g., Moran & Carroll, 1994) in which viewpoints adopted by actors from one speciality (speciality specific-viewpoints) or by the whole team (integrated viewpoints) are not explicitly represented. It means documenting the designs according to the various field-dependent constraints used to assess them as well as the combination and the weighting of these constraints as they are negotiated by the team.

3   Conclusion and perspectives

Our socio-cognitive approach has enabled us to construct a preliminary typology of reuse situations with their specific processes and their specific needs for supporting them. On the basis of results of empirical studies we have constructed a classification of reuse situations characterised by distinctive processes involved in reuse. This classification goes beyond those which are based on the type of the component reused. Three dimensions have been identified : prospective versus retrospective reuse; reuse in planning versus reuse in translating; reuse as a problem solving mechanism versus reuse as an argumentation mechanism. This cognitive classification has allowed us to identify which cognitive mechanisms are specific to certain reuse situations and which support is needed depending on the reuse situation. Further work should be done to identify other dimensions and refine this typology. This typology is now awaiting empirical use and validation.

**References**

Beguin, P. (1997) L'activité de travail : facteur d'intégration durant les processus de conception. In P. Bossard, C. Chanchevrier, et P. Leclair, (Eds): *Ingénierie concourante de la technique au social*. Economica. Paris

Bereiter, C., Burtis, P.J., & Scardamalia, M. (1988) Cognitive operations in constructing main points in written composition. *Journal of Memory and Language, 27*, 261-278.

Buratto, F. (2000) *Prescriptions des méthodes fonctionnelles et activité collective de conception. Cas de la conception de processus dynamiques.* Thèse d'Ergonomie, Université Paul Sabatier de Toulouse, 5 octobre 2000.

Burkhardt, J-M., & Détienne, F. (1995) An empirical study of software reuse by experts in object-oriented design. In K. Nordby, P. H. Helmersen, D. J. Gilmore and S. A. Arnesen (Eds): *Proceedings of INTERACT'95*. Chapman & Hall. 133-138.

Clement, J. (1988) Observed Methods for Generating Analogies in Scientific Problem Solving. *Cognitive Science, 12*, 563-586.

Détienne, F. (1991) Reasoning from a schema and from an analog in software code reuse. In J. Koenemann-Belliveau, T.G. Moher and S.P. Robertson (Eds): *Empirical studies of programmers, Fourth Workshop*. Ablex, Norwood, NJ. 5-22.

Détienne, F. (2002) *Software design: cognitive aspects*. Springer Verlag, Practitioners series.

Détienne, F., & Burkhardt, J.-M (2001). Des aspects d'ergonomie cognitive dans la réutilisation en génie logiciel. *Techniques et Sciences Informatiques, 20 (4),* 461-487.

Détienne, F., Rouet, J-F., Burkhardt, J-M., & Deleuze-Dordron, C. (1996) Reusing processes and documenting processes: toward an integrated framework. In T. R. G. Green, J. J. Canas, & C. P. Warren (Eds): *Proceedings of the Eight Conference on Cognitive Ergonomics*. p 139-144. (ECCE8, Granada, Spain, September 10-13.

De Terssac, G., & Chabaud, C. (1990) Référentiel opératif commun et fiabilité. In J. Leplat et G. de Terssac (Eds): *Les facteurs humains de la fiabilité dans les systèmes complexes*. Paris, Octarès.

De Vries, E. (1993) The role of case-based reasoning in architectural design : Stretching the design problem space. In W. Visser (Ed.): *Proceedings of the Workshop of the Thirteenth International Joint Conference on Artificial Intelligence "Reuse of designs : an interdisciplinary cognitive approach"*. Chambery August 29, 1993: INRIA Rocquencourt. B1-B13.

De Vries, E (1994) *Structuring information for design problem solving.* PhD Thesis. Eindhoven University, NL.

Eastman, C. M. (1969) Cognitive processes and ill-defined problems: a case study from design. In D.E. Walker and L. M. Norton (Eds): *Proceedings of the First Joint International Conference on Artificial Intelligence*. Bedford, MA: MITRE.

Falzon, P.(1994) Dialogues fonctionnels et activité collective. *Le Travail Humain, 57(4),* 299-312.

Falzon, P., & Darses, F. (1996) Collective Design Process. COOP Group (Ed): *Second International Conference on the Design Cooperative Systems*. June 12-14, 1996, Juan les pins, France.

Falzon, P., Bisseret, A., Bonnardel, N., Darses, F., Détienne, F., & Visser, W. (1990) Les activités de conception: l'approche de l'ergonomie cognitive. *Actes du Colloque Recherches sur le design: Incitations, implications, interactions*. Compiègne, 17-19 octobre 1990.

Flower, L., & Hayes, J.R. (1981). Plans that guide the composing process. In C. Frederiksen and J. Dominic (Eds.): *Writing: the nature, development and teaching of written communication. Vol. 2: writing: process, development and communication*. Hillsdale, NJ: Erlbaum. 39-58.

Gentner, D. (1989) The mechanisms of analogical learning. In S. Vosniadou & A. Ortny (Eds): Similarity and Analogical Reasoning. New York: Cambridge University Press. P 199-241.

Gray, W.D., & Anderson, J. R. (1987) Change-episode in coding: when and how programmers change their code?  In G. M. Olson, S. Sheppard and E. Soloway (Eds): *Empirical studies if programmers, second workshop*. Ablex. 185-197.

Grinter, R. (1999) Systems Architecture: Product Designing and Social Engineering. *WACC'99*, San Francisco, CA, USA.

Guindon, R. (1990) Designing the design process: exploiting opportunistic thoughts. *Human-Computer Interaction, 5*, 305-344.

Guindon, R. (1992) Requirements and design of Design Vision, an object-oriented graphical interface to an intelligent software design assistant. *Proceedings of CHI'92*. ACM Press, 499-506.

Hayes, J.R., & Flower, L. (1980). Identifying the organization of writing processes. In L.W. Gregg & E. Steinberg (Eds.): *Cognitive processes in writing*. Hillsdale, NJ: Erlbaum. 3-30.

Karsenty, L. (1996) An empirical evaluation of design rationale documents. In M.J. Tauber, V. Belloti, R. Jeffries, J.D. Mackinlay and J. Nielsen (Eds): *CHI'96 conference proceedings.* Addison Wesley. 150-156.

Lange, B.M., & Moher T.G. (1989) Some strategies of reuse in an object-oriented programming environment. In K. Bice and C. Lewis (Eds.): *Proceedings of CHI'89 Conference on Human Factors in Computing Systems*. ACM Press. 69-73.

Maiden, N. (1991) Analogy as a paradigm for specification reuse. Software Engineering Journal, 3-15.

Martin, G., Détienne, F., & Lavigne, E. (2000a) An ergonomic study on the design process in Concurrent Engineering. *Workshop3, COOP'2000*. 23 mai 2000, Sophia Antipolis, France.

Martin, G., Détienne, F., & Lavigne, E. (2000b). Negotiation in collaborative assessment of design solutions: an empirical study on a Concurrent Engineering process. *CE'2000, International Conference on Concurrent Engineering*. Lyon, France, 17-20 juillet 2000.

Martin, G., Détienne, F., & Lavigne, E. (2001). Analysing viewpoints in design through the argumentation process. *INTERACT'2001*. Tokyo, Japan, July 9-13.

Moran, T. P., & Carroll, J. M., Eds (1994) *Design Rationale: Concepts, Techniques, and Use*. Hillsdale, NJ: Erlbaum.

Perelman, C., & Olbrechts-Tyteca, L. (1992) *Traité de l'argumentation*. Edition de l'université de Bruxelles.

Plantin, C. ( 1996). *L'argumentation*. Seuil.

Rasmussen, J., & Lind, M. (1982) *A model of human decision making in complex systems and its use for design of sytem control strategies*. Roskilde, Danemark, RISO, M-2349.

Rosson, M.B., & Carroll, J.M. (1993) Active programming strategies in reuse. *Proceedings of ECOOP'93, Object-Oriented Programming*. Berlin: Springer-Verlag. 4-18 .

Rouet, J-F., Deleuze-Dordron, C., & Bisseret, A. (1995a) Documentation as part of design: exploratory field studies. In K. Nordby, P. H. Helmersen, D. J. Gilmore and S. A. Arnesen (Eds): *Proceedings of INTERACT'95*. Chapman & Hall. 213-216.

Sutcliffe, A., & Maiden, N. (1991). Analogical software reuse : empirical investigations of analogy-based reuse and engineering practices. *Acta Psychologica, 78*, 173-197.

Simon, H.A. (1973) The structure of ill-structured problems. *Artificial Intelligence, 4*, 181-201.

van Dijk, T.A., & Kintsch, W. (1983) *Strategies of Discourse Comprehension*. New York: Academic Press.

Visser, W. (1987) Strategies in Programming Programmable Controllers: A Field Study on a Professional Programmer. In G. M. Olson, S. Sheppard and E. Soloway (Eds): *Empirical Studies of programmers: second workshop*. Ablex. p 217-230.

Visser, W. (1995). Use of episodic knowledge and information in design problem solving. *Design Studies, 16(2),* 171-187.

Visser, W., & Hoc, J.-M. (1990). Expert software design strategies. In J.-M. Hoc, T. R. G. Green, R. Samurçay, & D. Gilmore (Eds.): *Psychology of Programming*. Academic Press. 235-250.

Visser, W., & Trousse, B. (1993). Reuse of designs: desperately seeking an interdisciplinary cognitive approach. In W. Visser (Ed.): *Workshop of the 13th International Joint Conference on Artificial Intelligence "Reuse of designs : an interdisciplinary cognitive approach"*. Chambery August 29, 1993: INRIA. 1-14.